\documentclass{llncs}

\usepackage{amstext,amssymb,amsfonts,latexsym} 
\usepackage{pifont}

 \newcommand{\bs}{\bigskip}
 \newcommand{\ms}{\medskip}
 
 \newcommand{\s}{\smallskip}
 \newcommand{\hs}[1]{\hspace*{ #1 mm}}

 \newcommand{\nat}{\mathbb{N}}
 \newcommand{\integers}{\mathbb{Z}}
 
 \newcommand{\complex}{\mathbb{C}}

 \newcommand{\vsigma}{\mbox{\boldmath $\sigma$}}
 \newcommand{\vtau}{\mbox{\boldmath $\tau$}}
 \newcommand{\vepsilon}{\mbox{\boldmath $\epsilon$}}
 \newcommand{\vd}{\mbox{\boldmath $d$}}
 \newcommand{\vp}{\mbox{\boldmath $p$}}
 \newcommand{\vq}{\mbox{\boldmath $q$}}
 
 \newcommand{\svtau}{\mbox{\boldmath ${}_{\tau}$}}
 \newcommand{\svepsilon}{\mbox{\boldmath ${}_{\epsilon}$}}
 \newcommand{\svd}{\mbox{\boldmath ${}_{d}$}}
 \newcommand{\svx}{\mbox{\boldmath ${}_{x}$}}
 \newcommand{\vx}{\mbox{\boldmath $x$}}
 \newcommand{\vN}{\mbox{\boldmath $N$}}
 
 \newcommand{\vR}{\mbox{\boldmath $R$}}

 \newcommand{\ie}{\textrm{i.e.},\hspace*{2mm}}
 \newcommand{\eg}{\textrm{e.g.},\hspace*{2mm}}

 \def\bbox{\vrule height6pt width6pt depth1pt}

 \newenvironment{proofqed}{\par \noindent 
            {\it Proof. \hs{2}}}{\hfill\bbox \vspace*{3mm}}
 
 \newenvironment{proofof}[1]{\vspace*{5mm} \par \noindent 
         {\it Proof of #1.\hs{2}}}{\hfill\bbox \vspace*{3mm}}

 \newcommand{\floors}[1]{\lfloor #1 \rfloor}

 \newcommand{\runtime}{\mathrm{Time}}

 \newcommand{\qbit}[1]{| #1 \rangle}
 \newcommand{\bra}[1]{\langle #1 |}
 \newcommand{\ket}[1]{| #1 \rangle}
 \newcommand{\measure}[2]{\langle #1 | #2 \rangle}

\begin{document} 
\begin{flushright}
{\tiny To appear in the Proceedings \\ of the 24th International
Symposium \\ on Mathematical Foundations \\ of Computer Science,
September, 1999 \\}
\end{flushright}
\ms

\begin{center}
{\large A Foundation of \\ Programming a Multi-Tape Quantum Turing
Machine} \s\\
{\small {\it (Preliminary Version)}} \ms\\ 
Tomoyuki Yamakami \footnote{This work is partly supported by NSERC
 Postdoctoral Fellowship and DIMACS Fellowship.} \ms\\ 
{\small Department of Computer Science \\ Princeton University \\
35 Olden Street, Princeton, NJ 08544}
\end{center}

\begin{abstract}
The notion of quantum Turing machines is a basis of quantum complexity
theory. We discuss a general model of multi-tape, multi-head Quantum
Turing machines with multi final states that also allow tape heads to
stay still.
\end{abstract}

\section{Introduction}

A quantum Turing machine (QTM) is a theoretical model of quantum
computers, which expands the classical model of a Turing machine (TM)
by allowing quantum interference to take place on their computation
paths.  Designing a QTM in general, however, is significantly harder
than that of a classical TM because of its {\em well-formedness}
condition as well as its halting condition, known as the {\em timing
problem}. Recently Bernstein and Vazirani \cite{BV97} initiated a
study of quantum complexity theory founded on a restrictive model: a
one-head, multi-track, stationary, dynamic, normal form,
unidirectional QTM (for definitions, see Section \ref{sec:basics})
that prohibits a tape head to stay still. We call such a restrictive
QTM {\em conservative} for convenience.

One may find easier to program a less restrictive QTM when he wishes
to solve a problem on a quantum computer. In this paper we wish to
introduce a QTM as general as possible. In Section \ref{sec:basics},
we introduce a multi-tape, multi-head QTM with multi final states that
also allows tape heads to stay still.  Although many variations of
QTMs are known to be polynomially equivalent \cite{BV97,Yao93},
unsolved is the question of what is the degree of polynomials of these
simulation overhead.  As we will show in Section \ref{sec:simulation},
any multi-tape, multi-head, well-formed QTM can be effectively
simulated by a conservative QTM with only cubic polynomial slowdown.

Our primary goal is to contribute to the foundation of programming a
handy QTM. In Section \ref{sec:lemmas}, we will prove two fundamental
lemmas: Well-formedness Lemma and Completion Lemma, which are
important tools in constructing a QTM. The lemmas expand the results
of Bernstein and Vazirani \cite{BV97}, who considered mostly
conservative QTMs. Using the lemmas, we will show that any computation
of a well-formed QTM can be reversed on a well-formed QTM with
quadratic polynomial slowdown.  We will also address the timing
problem in Section \ref{sec:simulation}.  In Section \ref{sec:oracle},
we will focus on an oracle QTM with multi query tapes and multi
oracles. For any oracle QTM $M$, we can build an oracle QTM, similar
to the classical case, that simulates $M$ with a fixed number of
queries of fixed length on every computation path.


\section{Definition of Quantum Turing Machines}\label{sec:basics}

This section briefly describes the formal definition of quantum Turing
machines.  For our purpose, we wish to make the definition as general
as possible. Here we present a definition that is slightly more
general than the one given in {\cite{BV97,BBBV97}}.

A $k$-tape {\em quantum Turing machine} (QTM) $M$ is a quintuple
$(Q,\{q_0\},Q_f,\Sigma_1\times\Sigma_2\times\cdots\times\Sigma_k,\delta)$,
where each $\Sigma_i$ is a finite alphabet with a distinguished blank
symbol $\#$, $Q$ is a finite set of internal states including an
initial state $q_0$ and $Q_f=\{q^1_f,q^2_f,\ldots,q^m_k\}$, a set of
final states, and $\delta$ is a multi-valued, {\em quantum transition
function} from
$Q\times\Sigma_1\times\Sigma_2\times\cdots\times\Sigma_k$ to
$\complex^{Q\times\Sigma_1\times\Sigma_2 \times\cdots\times\Sigma_k
\times\{R,N,L\}^k}$. (Note that $\delta(q_f^i,\vsigma)$ must be
defined.)  For brevity, write $\tilde{\Sigma}^{(k)}$ for
$\Sigma_1\times\cdots\times\Sigma_k$. A QTM has two-way infinite tapes
of cells indexed by $\integers$ and read/write tape heads that move
along the tapes. Directions $R$ and $L$ mean that a head steps right
and left, respectively, and direction $N$ mean that a head makes no
movement.  We say that all tape heads move {\em concurrently} if they
move in the same direction at any time (in this case, \eg we write
$\delta(p,\vsigma,q,\vtau,d)$ instead of
$\delta(p,\vsigma,q,\vtau,\vd)$). We call a QTM {\em dynamic} if its
heads never stay still. A QTM is {\em unidirectional} if, for any
$p_1,p_2,q\in Q$, $\vsigma_1,\vsigma_2\in\tilde{\Sigma}^{(k)}$, and
$\vd_1,\vd_2\in\{L,N,R\}^k$,
$\delta(p_1,\vsigma_2,q,\vtau_1,\vd_1)\cdot
\delta(p_2,\vsigma_2,q,\vtau_2,\vd_2)\neq0$ implies $\vd_1=\vd_2$.

We assume the reader's familiarity with the following terminology: a
{\em time-evolution operator}, a {\em configuration} and {\em final
configuration}, a {\em superposition} and a {\em final superposition},
a {\em well-behaved} and {\em stationary} QTM, and the {\em acceptance
probability} of a QTM. For their definitions, see \cite{BV97}. 

Here are ones different from \cite{BV97}.  A QTM is in {\em normal
form} if, for every $i\in\{1,2,\ldots,m\}$, there exists a direction
$\vd_i\in\{L,N,R\}^k$ such that $\delta(q_f,\vsigma)=
\qbit{q_0}\qbit{\vsigma}\qbit{\vd_i}$ for any
$\vsigma\in\tilde{\Sigma}^{(k)}$. A QTM $M$ is called {\em
synchronous} if, for every $\vx$, any two computation paths of $M$ on
$\vx$ reach final configurations at the same time.  The {\em running
time} of $M$ on $\vx$ is defined to be the minimal number $T$ such
that, at time $T$, all computation paths of $M$ on $\vx$ reach final
configurations. We write $\runtime_M(\vx)$ to denote the running time
of $M$ on $\vx$ if one exists; otherwise, it is undefined. We say that
$M$ on input $\vx$ {\em halts in time $T$} if $\runtime_M(\vx)$ exists
and $\runtime_M(\vx)=T$.  A QTM is {\em well-formed} if its
time-evolution operator preserves the $L_2$-norm.  A multi-tape QTM is
said to be {\em conservative} if it is a well-formed, stationary,
dynamic, unidirectional QTM in normal form with concurrent head move.
We write $\mu_M(\vx)$ to denote the probability that $M$ accepts input
$\vx$.

Throughout this paper, $T$ denotes a function from $\Sigma^*$ to
$\nat$.


\section{Fundamentals of Quantum Turing Machines}
\label{sec:lemmas}

In this section, we will prove two lemmas that are essential tools in
programming a well-formed QTM: {\em Well-Formedness Lemma} and {\em
Completion Lemma}.

For convenience, the head move directions $R$, $N$, and $L$ are
identified with $-1$, $0$, and $+1$, respectively.

\paragraph{\bf Well-Formedness Lemma.}

One of the most significant feature of a QTM is the well-formedness
condition on its quantum transition function that reflects the
unitarity of their corresponding time-evolution operators.  Here we
present in Lemma \ref{lemma:general-formedness} three local
requirements for a quantum transition function whose associated QTM is
well-formed.

Let $M=(Q,\{q_0\},Q_f,\Sigma_1\times\cdots\times\Sigma_k,\delta)$ be a
$k$-tape QTM.  Recall that $\tilde{\Sigma}^{(k)}$ stands for
$\Sigma_1\times\cdots\times\Sigma_k$. We introduce the notation
$\delta[p,\vsigma,\vtau|\vepsilon]$.  Let $D=\{0,\pm1\}$,
$E=\{0,\pm1,\pm2\}$, and $H=\{0,\pm1,\natural\}$.  Let
$(p,\vsigma,\vtau)\in Q\times (\tilde{\Sigma}^{(k)})^2$ and
$\vepsilon\in E^k$. Define $D_{\svepsilon} =\{\vd\in D^k\mid \forall
i\in\{1,\ldots,k\}(|2d_i-\epsilon_i|\leq1)\}$ and $E_{\svd}
=\{\epsilon\in E^k\mid \vd\in D_{\svepsilon}\}$, where
$\vd=(d_i)_{1\leq i\leq k}$ and $\vepsilon=(\epsilon_i)_{1\leq i\leq
k}$.  Let $h_{\svd,\svepsilon} =(h_{d_i,\epsilon_i})_{1\leq i\leq k}$,
where $h_{d,\epsilon}=2d-\epsilon$ if $\epsilon\neq 0$ and
$h_{d,\epsilon}=\natural$ otherwise. Finally, we define
$\delta[p,\vsigma,\vtau|\vepsilon]$ as follows:
$\delta[p,\vsigma,\vtau|\vepsilon] = \sum_{q\in Q} \sum_{\svd\in
D_{\svepsilon}} \delta(p,\vsigma,q,\vtau,\vd)|E_{\svd}|^{-1/2}
\qbit{q} \qbit{h_{\svd,\svepsilon}}$.

\begin{lemma}\label{lemma:general-formedness}
(Well-Formedness Lemma)\hs{2} A $k$-tape QTM
$M=(Q,\{q_0\},Q_f,\Sigma_1\times\cdots\times\Sigma_k,\delta)$ is
well-formed iff the following three conditions hold.
\begin{enumerate}
\item (unit length) $\|\delta(p,\vsigma)\|=1$ for
all $(p,\vsigma)\in Q\times \tilde{\Sigma}^{(k)}$.

\item (orthogonality) $\delta(p_1,\vsigma_1)
\cdot\delta(p_2,\vsigma_2)=0$ for any distinct pair $(p_1,\vsigma_1),
(p_2,\vsigma_2)\in Q\times \tilde{\Sigma}^{(k)}$.

\item (separability) $\delta[p_1,\vsigma_1,\vtau_1|\vepsilon]\cdot
\delta[p_2,\vsigma_2,\vtau_2|\vepsilon']=0$ for any distinct pair
$\vepsilon,\vepsilon'\in E^k$ and for any pair
$(p_1,\vsigma_1,\vtau_1), (p_2,\vsigma_2,\vtau_2)\in Q\times
(\tilde{\Sigma}^{(k)})^2$.
\end{enumerate}
\end{lemma}

The proof of the lemma is similar to that of Theorem 5.3 in
\cite{BV97}. Note that, since any two distinct tapes do not interfere,
a $k$-tape QTM must satisfy the $k$ independent conditions for the
case $k=1$. We leave the detail to the reader.

\paragraph{\bf Completion Lemma.}\label{sec:completion}

A quintuple
$M=(Q,\{q_0\},Q_f,\Sigma_1\times\cdots\times\Sigma_k,\delta)$ is
called a {\em partial} QTM if $\delta$ is a partial quantum transition
function that is defined on a subset $S$ of
$Q\times\Sigma_1\times\cdots\times\Sigma_k$. If $\delta$ satisfies the
three conditions of Lemma \ref{lemma:general-formedness} on all
entries of $\delta$, then we call $M$ a {\em well-formed partial} QTM
\cite{BV97}.

Completion Lemma says that any well-formed partial QTM can be expanded
to a well-formed QTM.

\begin{lemma}\label{lemma:completion}(Completion Lemma)\hs{2} 
For every $k$-tape, well-formed partial QTM with quantum transition
function $\delta$, there exists a $k$-tape, well-formed QTM with the
same state set and alphabet whose transition function $\delta'$ agrees
with $\delta$ whenever $\delta$ is defined.
\end{lemma}

To show the lemma, we first consider how to change the basis of a
given QTM. Let
$M=(Q,\{q_0\},Q_f,\Sigma_1\times\cdots\times\Sigma_k,\delta)$ be a
given QTM. We first partition $\complex^{Q\times H^k}$ into mutually
orthogonal spaces $\{\complex_{\svepsilon}\mid \vepsilon\in E^k\}$
such that (i) $\complex^{Q\times H^k}= span\{\complex_{\svepsilon}\mid
\vepsilon\in E^k\}$ and (ii) for any $\vepsilon\in E^k$ and any
$(p,\vsigma,\vtau)\in Q\times (\tilde{\Sigma}^{(k)})^2$,
$\delta[p,\vsigma,\vtau|\vepsilon]\in \complex_{\svepsilon}$. Note
that if $|\vepsilon|\neq|\vepsilon'|$ then $\complex_{\vepsilon}\cap
\complex_{\vepsilon'}=\emptyset$.  For each $\vepsilon\in E^k$, let
$B_{\svepsilon}$ be an orthonormal basis for
$\complex_{\svepsilon}$. Let $B$ be the union of all such
$B_{\svepsilon}$'s.

We assume that, at time $t$, $M$ in state $p$ scans symbol $\vsigma$
and that $\delta$ maps $(p,\vsigma)$ to$\sum_{q,\svtau,\svd}
\delta(p,\vsigma,q,\vtau,\vd) \qbit{q}\qbit{\vtau}\qbit{\vd}$.  Define
the change of basis from $Q\times\{L,N,R\}^k$ to $B\times E^k$ by
mapping $\qbit{q}\qbit{\vd}$ into $\sum_{w\in B}\sum_{\svepsilon\in
E_{\svd}} \measure{h_{\svd,\svepsilon},q}{w} |E_{\svd}|^{-1/2}
\qbit{w}\qbit{\epsilon}$. Let $U_1$ denote this transform.  This
matrix $U_1$ is unitary because $\bra{\vd,q}U_1^*U_1\ket{q',\vd'} =
\sum_{\svepsilon\in E_{\svd}\cap E_{\svd'}}
\measure{h_{\svd',\svepsilon},q'}{q,h_{\svd,\svepsilon}}
(|E_{\svd}|\cdot|E_{\svd'}|)^{-1/2} = \measure{\vd,q}{q',\vd'}$, which
implies \linebreak 
$U_1^*U_1=I$. It is known in \cite{BV97} that $U_1$ preserves the
$L_2$-norm iff $U_1$ is unitary.

Let $\delta'(p,\vsigma)$ denote $U_1 \delta(p,\vsigma)$ for any
$(p,\vsigma)\in S$.  In what follows, we show that $\delta'$ is
``unidirectional'' in the sense that if
$\delta'(p,\vsigma,v,\vtau,\vepsilon)\cdot
\delta'(p',\vsigma',v,\vtau',\vepsilon')\neq0$ then
$\vepsilon=\vepsilon'$.  Let $\vepsilon$ and $\vepsilon'$ be distinct
and in $E^k$.  Note that the separability condition ensures that
$\delta[p,\vsigma,\vtau|\vepsilon]\cdot
\delta[p',\vsigma',\vtau'|\vepsilon']=0$ for any
$(p',\vsigma',\vtau')\in Q\times(\tilde{\Sigma}^{(k)})^2$. Since
$\delta[p,\vsigma,\vtau|\vepsilon] = \sum_{v\in B}
\delta'(p,\vsigma,v,\vtau,\vepsilon) \qbit{v}\in
\complex_{\svepsilon}$, $\delta'(p,\vsigma,v,\vtau,\vepsilon)=0$ for
any $v\in B_{\svepsilon'}$ if $\vepsilon\neq \vepsilon'$. Therefore,
$\delta'$ is ``unidirectional.''

The transform $U_1$ is useful to show Completion Lemma.  We go back to
the formal proof of Completion Lemma.

\begin{proofof}{Lemma \ref{lemma:completion}}
Let $M=(Q,\{q_0\},Q_f,\Sigma_1\times\cdots\times\Sigma_k,\delta)$ be a
given QTM.  Let $U_1$ be defined as above.  As shown above, $U_1$ is
unitary.  As a result, $\delta'(S)$ is a set of orthonormal vectors
since so is $\delta(S)$.

For each $v\in B$, let $\vepsilon_{v}$ be $\vepsilon$ such that
$\delta'(p,\vsigma,v,\vtau,\vepsilon)\neq0$ for some
$(p,\vsigma,\vtau)\in Q\times (\tilde{\Sigma}^{(k)})^2$ if any, and
let $\vepsilon_{v}=(1)_{1\leq i\leq k}$ otherwise. Since $\delta'$ is
``unidirectional'', $\vepsilon_{v}$ is uniquely determined.  This
implies that we can define the vector $\delta''(p,\vsigma)$ as
$\delta''(p,\vsigma) = \sum_{v\in B}
\sum_{\svtau\in\tilde{\Sigma}^{(k)}}
\delta'(p,\vsigma,v,\vtau,\epsilon_{v}) \qbit{v}\qbit{\tau}$.

Now we expand $\delta''$ to $Q\times\tilde{\Sigma}^{(k)}$ by adding
arbitrarily extra orthonormal vectors associated with elements in
$Q\times\tilde{\Sigma}^{(k)} - S$. Let $\overline{\delta}''$ be such
an expansion of $\delta''$. We define $\overline{\delta}'$ by
$\overline{\delta}'(p,\vsigma) = \sum_{v\in B}
\sum_{\svtau\in\tilde{\Sigma}^{(k)}} \sum_{\svepsilon\in E^k}
\overline{\delta}''(p,\vsigma,v,\vtau,\vepsilon)
\qbit{v}\qbit{\vtau}\qbit{\vepsilon}$.

We then apply the inverse transform $U_1^*$ to
$\overline{\delta}'(Q\times\tilde{\Sigma}^{(k)})$ and let
$\overline{\delta}$ be the result obtained. Define $\overline{M}=
(Q,\{q_0\},Q_f,\tilde{\Sigma}^{(k)},\overline{\delta})$. Since $U_1$
is unitary, $\overline{M}$ must be well-formed.
\end{proofof}

Completion Lemma also enables us to use a $k$-tuple of a single
alphabet, $\Sigma^k$, instead of
$\Sigma_1\times\ldots\times\Sigma_k$. In the following sections, we
will deal only with a $k$-tape QTM with tape alphabets $\Sigma^k$.

\section{Simulation of Quantum Turing Machines}\label{sec:simulation}

In this section we demonstrate several simulation results using the
main lemmas in Section \ref{sec:lemmas}.  Since we are interested only
in the acceptance probability of a QTM, the ``simulation'' of a QTM
$M$ by another QTM $M'$ in this paper regards with the statement that
$N$ produces the same acceptance probability as $M$ does. More
formally, we say that $M'$ {\em simulates $M$ with slowdown $f$} if,
for every $\vx$, $\mu_{M'}(\vx)=\mu_M(\vx)$ and $\runtime_{M'}(\vx)=
f(\runtime_{M}(\vx))$.

Assume that $M$ is a $k$-tape well-formed QTM running in time $T(\vx)$
on input $\vx$. For $m\geq1$, let $\tilde{M}_{T,m}$ denote the
$(k+m)$-tape QTM that, on input $\vx$ in tapes 1 to $k$ and
$1^{T(\vx)}$ in tape $k+1$ and empty elsewhere, behaves like $M$ on
input $\vx$ except that the heads in tapes $k+1$ to $k+m$ idle in the
start cells. The tape alphabets of $\tilde{M}_{T,m}$ for tapes 1 to
$k$ are the same as $M$'s.

For convenience, let $M(\qbit{\phi})$ denote the final superposition
of $M$ that starts with superposition $\qbit{\phi}$. In the case where
$\qbit{\phi}$ is an initial configuration with input $\vx$, we write
$M(\qbit{\vx})$ for $M(\qbit{\phi})$.

For any pair $\vsigma=(\sigma_i)_{1\leq i\leq k}$ and
$\vtau=(\tau_j)_{1\leq j\leq m}$, $\vsigma*\vtau$ denotes the
$(k+m)$-tuple $(\sigma_1,\ldots,\sigma_k,\tau_1,\ldots,\tau_m)$. In
particular, we write $s*\vsigma$ for $(s)*\vsigma$ and $\vsigma*s$ for
$\vsigma*(s)$.

\paragraph{\bf Simulation by Synchronous Machines.}

We show how to transform any well-formed QTM into a well-formed,
synchronous QTM with a single final state with the help of the
information on its running time.

\begin{lemma}\label{lemma:final-state}
Let $M$ be a $k$-tape, well-formed QTM that halts in time $T(\vx)$ on
any $k$-tuple input string $\vx$. Then, there exists a $(k+2)$-tape,
well-formed, synchronous QTM $M'$ with a single final state such that,
on input $(\vx,1^{T(\vx)})$, it halts in time $2T(\vx)+2$, the last
two tape heads move back to the start cells, leaving $1^{T(\vx)}$
unchanged, and $\mu_{M'}(\vx,1^{T(\vx)})=\mu_M(\vx)$. If $M$ already
has a single final state, then $M'$ needs only $k+1$ tapes and
satisfies $M'(\qbit{\vx,1^{T(\vx)}})=
\tilde{M}_{T,1}(\qbit{\vx,1^{T(\vx)}})$.
\end{lemma}

\begin{proofqed}
Let $M=(Q,\{q_0\},Q_f,\Sigma^k,\delta)$ be a given QTM with
$Q_f=\{q_f^1,q_f^2,\ldots,q_f^m\}$.  By Completeness Lemma, it
suffices to build a partial QTM $M'$ that satisfies the lemma.  Assume
that $\vx$ is given in tapes $1$ to $k$ and $1^{T(\vx)}$ is in tape
$k+1$. Tape $k+2$ is initially empty. The QTM $M'$ simulates each step
of the computation of $M$ using tapes $1$ to $k$, together with
stepping right in tape $k+1$, which counts the number of steps
executed by $M$. When $M'$ arrives at any final configuration of $M$
with final state $q_f^i$, $1\leq i\leq m$, at time exactly $T(\vx)$,
$M'$ deposits the number $i$ (as a single tape symbol) onto tape
$k+2$, freeing itself from state $q_f^i$. Then, $M'$ moves its $k+1$st
tape head back to the start cell in $T(\vx)+2$ steps and enters its
own final state $q_f$. Thus, the running time of $M'$ is exactly
$2T(\vx)+2$.

It is not difficult to check the well-formedness of $M'$ using
Well-Formedness Lemma. Note that the acceptance probability of $M$
does not change during the above simulation process. Thus,
$\mu_{M'}(1^{T(\vx)},\vx)=\mu_M(\vx)$.

If $M$ already has a single final state $q_f$, we modify the above
procedure in the following fashion. Firstly, we replace every
occurrence of $q_f$ in $\delta$ by $\hat{q}_f$. Secondly, we apply the
above simulation procedure. Thirdly, after the simulation, we force
$M'$ to enter $q_f$ as its final state exactly when the $k+1$st tape
head returns to the start cell. In this case, we do not need the
$k+2$nd tape at all.
\end{proofqed}

\paragraph{\bf Simulation by Machines with Concurrent Head Move.}

The simulation of a multi-tape QTM by a single tape QTM is a central
subject in this subsection. We show that any multi-tape, well-formed
QTM can be simulated by a certain well-formed, well-behaved QTM with
concurrent head move. The simulation overhead here is a quadratic
polynomial. This result makes it possible to simulate a multi-tape QTM
by a single tape QTM with quadratic polynomial slowdown.

\begin{proposition}\label{prop:concurrent}
Let $M$ be a $k$-tape, well-formed QTM that halts in time $T(\vx)$ on
input $\vx$. There exists a $(k+2)$-tape, well-formed, well-behaved
QTM $M'$ with concurrent head move such that $M'$, on input $\vx$ in
tapes 1 to $k$ and empty elsewhere, simulates $M$ in time
$2T(\vx)^2+(2k+9)T(\vx)+4$.  Moreover, if $M$ is synchronous, dynamic,
unidirectional, or normal form, so is $M'$. In particular, when $M$ is
synchronous, $M'$ can be made stationary with extra $T(\vx)+1$ steps.
\end{proposition}

\begin{proofqed}
Given a QTM $M$, we construct a new QTM $M'$ that simulates in
$4r+2k+7$ steps the $r$th step of $M$ by moving its heads back and
forth in all tapes concurrently and by expanding the simulation area
by 2.  Thus, the new QTM needs $\sum_{r=1}^{T(\vx)}(4r+2k+7)$ steps
(with an additional pre-computation of 4 steps) to complete this
simulation on input $\vx$.

Let $M=(Q,\{q_0\},Q_f,\Sigma^k,\delta)$ be a given QTM.  The desired
$M'$, starting from state $\hat{q}_0$, works as follows. Initially, in
four steps we mark \$ in the start cell in tape $k+2$ and we set up
the simulation area of three cells (which are indexed $-1$, $0$, $1$)
in tape $k+1$, each of which holds the record of the head position of
$M$. We will maintain this record in tape $k+1$ by updating a symbol
$(\sigma_i)_{1\leq i\leq k}$ in each cell, where $\sigma_i=1$ means
that the $i$th tape head rests in the current cell. Finally, $M'$
enters state $(q_0,\tau_0,\vd_0)$, where $\vtau_0=\vd_0=(\$)_{1\leq
i\leq k}$.

At round $r$, $1\leq r\leq T(\vx)$, we simulate the $r$th step of the
computation of $M$ in $4r+2k+7$ steps. We start with state
$(p,\vtau_0,\vd_0)$, provided that $p$ is a current state of
$M$. Moving the head rightward along all tapes toward the end of the
simulation area, we collect the information on a $k$-tuple
$\vtau=(\tau_i)_{1\leq i\leq k}$ of tape symbols being scanned by $M$
at time $r$ and we then remember it by changing our internal state
from $(p,\vtau_0,\vd_0)$ to $(p,\vtau,\vd_0)$. After the head arrives
at the first blank cell, by applying the transition $\delta(p,\vtau)$,
we change $(p,\vtau,\vd_0)$ into $(q,\vtau,\vd)$ if
$\delta(p,\vtau,q,\vsigma,\vd)$ is non-zero. To end this simulation
phase, we update the head position marked in tape $k+1$ (by using
$\vd$) and tape symbols (by using $\vtau$) by moving the head leftward
to the first blank cell in tape $k+1$. Whenever the head reaches an
end of the simulation area, we expand this area by 1 by writing the
symbol $(0)_{1\leq i\leq k}$ in its boundary blank cell. After the
simulation phase, $M'$ enters state $(q,\vtau_0,\vd_0)$.
 
Suppose that $M$ is in normal form. It is easy to verify that no
well-formed QTM in normal form has more than two final state. Let
$q_f$ be a single final state of $M$. Adding the rule
$\delta'((q_f,\vtau_0,\vd_0),\vsigma)=
\qbit{\hat{q}_0}\qbit{\vsigma}\qbit{\vR}$ makes $M'$ be in normal form.
If $M'$ is synchronous, then $M'$ can use the marker \$ in tape $k+2$
to move its head back to the start cell and erase \$ from the tape in
$T(\vx)+1$ steps. This last movement forces $M'$ to be stationary.
\end{proofqed}

Any QTM with concurrent head move can reduce the number of tapes by
merging a $k$-tuple of tape symbols which the head is scanning, into a
single tape symbol.

\begin{lemma}\label{lemma:one-tape}
Let $1\leq m\leq k$. Let $M$ be a $k$-tape, well-formed QTM with
concurrent head move that, on input $\vx$ in tapes $1$ to $m$ and
empty elsewhere, halts in time $T(\vx)$. There exists an $m$-tape,
well-formed QTM $M'$ such that, on input $\vx$, halts in time $T(\vx)$
and simulates the computation of $M$ on $\vx$. If $M$ is dynamic,
synchronous, stationary, unidirectional, or normal form, so is $M'$.
\end{lemma}

\paragraph{\bf Simulation by Dynamic Machines.}\label{sec:dynamic}

This subsection is devoted to show that any well-formed QTM can be
simulated by a certain conservative QTM with quadratic polynomial
slowdown.

\begin{proposition}\label{prop:dynamic}
Let $M$ be a $k$-tape, well-formed, synchronous QTM that halts in time
$T(\vx)$ on input $\vx\in\Sigma^k$. There exists a $2k$-tape,
well-formed, stationary, synchronous, unidirectional, dynamic QTM $M'$
such that, on input $\vx$ in tape 1 to $k$ and empty elsewhere, $M'$
simulates $M$ in time $2T(\vx)^2+16T(\vx)+4$. If $M$ has a single
final state, then $M'$ is further in normal form.
\end{proposition}

\begin{proofqed}
The proof uses an idea of Yao \cite{Yao93}. Let
$M=(Q,\{q_0\},Q_f,\Sigma,\delta)$ be a given QTM with
$Q_f=\{q_f^1,q_f^2,\ldots,q_f^m\}$. We define the desired partial $M'$
so that it simulates the $r$ step of $M$ by a round of $4r+13$ steps
with all the heads moving concurrently. Since $M$ requires $T(\vx)$
steps, $M'$ needs $T(\vx)\sum_{r=1}^{T(\vx)}(4r+13)$ steps together
with a pre- and post-computation of $T(\vx)+4$ steps, which gives the
desired running time.

We first show the proposition for the special case $k=1$. Let
$x=x_1x_2\cdots x_m$ be an input given in tape 1. In the initial phase,
we create in four steps the configuration $(p_0,x_1'x_2\cdots
x_m,-1,\$11,-1)$, where $x_1'=(q_0,x_1)$ and $p_0$ is a distinguished
state of $M'$ and symbol \$ is in the cell indexed $-1$.

To understand the simulation phase, we associate a configuration $cf$
of $M$ with a certain configuration $cf'$ of $M'$ defined in the
following way. Assume that $cf=(q,cont,k)$, where $M$ in state $q$
scans symbol $\sigma$ in the cell indexed $k$ and $cont$ is the
content of the tape.  At the beginning of round $r$, $1\leq r\leq
T(\vx)$, we create the configuration $cf'$ of $M'$ which is of the
form $(p_0,cont'_k,-r,1^{r-1}\$1^{r+1},-r)$, where $1^{r-1}\$1^{r+1}$
is written in tape 2 with \$ in the cell indexed $-1$ (which marks the
simulation area) and $cont'_k$ is identical to $cont$ execpt that the
cell indexed $k$ has symbol $(q,\sigma)$ instead of $\sigma$.

To disregard any head direction that results from an application of
$\delta$, we treat as a single symbol the three consecutive symbols,
where the head of $M$ scans the middle symbol. In the course of the
simulation, we first search in tape $2$ the three consecutive symbols
$\sigma_0;(q,\sigma_1);\sigma_2$, where $M$ in state $q$ scans
$\sigma_1$, and encode them into the single symbol
$(\sigma_0,(q,\sigma_1),\sigma_2)$ by moving the head back and
forth. We then apply $\delta$ to this symbol with stepping right. This
makes $M'$ dynamic and also unidirectional. Finally, we decode the
result and update the content of tape $2$.

For each configuration at time $r$ of $M$ on input $x$, at the end of
the simulation, $M'$ produces its associated configuration. Therefore,
when $M$ enters a final configuration at time $T(x)$, $M'$ reaches a
configuration in which a tape symbol of the form $(q_f,\sigma)$ is
found in tape 2. When $M'$ finds such a symbol, it enters its own
final state $\hat{q}_f$ in exactly $T(\vx)$ steps. Let $\delta_1'$ be
the transition function for $M'$.

For a general case $k\geq1$, let $\vp=(p_j)_{1\leq j\leq k}$,
$\vsigma=(\sigma_j)_{1\leq j\leq k}$, and $\vtau=(\tau_j)_{1\leq j\leq
k}$. We first produce $\$;1;1$ in tapes $k+1$ to $2k$ and enters state
$\vp_0=(p_0)_{1\leq j\leq k}$ with changing symbol $\sigma_i$ in the
start cell in tape $i$ into $(q_0,\sigma_i)$. We then define
$\delta'_k(\vp,\vsigma*\vtau)$ to be the product
$\delta'_1(p_1,(\sigma_1,\tau_1)) \otimes
\delta'_1(p_2,(\sigma_2,\tau_2)) \otimes \cdots\otimes
\delta'_1(p_k,(\sigma_k,\tau_k))$.  Clearly, this QTM is well-formed,
stationary, and unidirectional. Note that the running time of the
$k$-tape QTM $M'$ does not depend on the number of tapes. Since $M$
halts at time $T(\vx)$, $M'$ finally enters state $(q_f^{i_j})_{1\leq
j\leq k}$ for some $k$-tuple $(i_j)_{1\leq j\leq k}$ at time
$2T(\vx)^2+16T(\vx)+4$.

In the case where $M$ has a single final state $q_f$, we can add the
new transition rule: $\delta'(\vq_f,\vsigma*\vtau) =
\qbit{\hat{q}_0}\qbit{\vsigma,\vtau}\qbit{\vR}$, where
$\vq_f=(q_f)_{1\leq j\leq k}$, which makes $M'$ be in normal form.
\end{proofqed}

Since the proposition regards with a unidirectional QTM, it also gives
an extension of Unidirection Lemma in \cite{BV97} to multi-tape QTMs.

Simply combining Propositions \ref{prop:concurrent} and
\ref{prop:dynamic} and Lemmas \ref{lemma:final-state} and
\ref{lemma:one-tape}, we obtain the following corollary.

\begin{corollary}
Let $M$ be a $k$-tape, well-formed QTM that, on input $x$ in tape 1
and empty elsewhere, runs in time $T(x)$. There exist a quartic
polynomial and a two-tape conservative QTM $M'$ such that, on input
$(1^{T(x)},x)$, $M'$ halts in time $p(T(x))$ and satisfies
$\mu_{M'}(1^{T(x)},x)= \mu_M(x)$.
\end{corollary}

Note that, by modifying the simulation given in the proof of
Proposition \ref{prop:concurrent} (with $O(S(\vx)T(\vx))$ slowdown,
where $S(\vx)$ is any space bound of $M$), we can achieve a much
tighter $O(T(x)^3)$ time bound. The detail is left to the reader.

\paragraph{\bf Reversing a Computation.}

First recall Definition 4.11 in \cite{BV97} that defines the notion:
$M_2$ {\em reverses the computation of} $M_1$. Different from
\cite{BV97}, we only assume that $M_1$ and $M_2$ are well-formed QTMs
(whose tape alphabets may differ) and that $M_1$ has a single final
state.  We show below that we can reverse the computation of any
well-formed QTM with quadratic polynomial slowdown.

\begin{theorem}\label{theorem:reverse}
Let $M$ be a $k$-tape, well-formed QTM with a single final state that
halts in time $T(\vx)$ on input $\vx$. There exist a quadratic
polynomial $p$ and a $2(k+1)$-tape, well-formed, synchronous, dynamic
QTM $M^R$ in normal form that, on input $\vx$ in tapes $1$ to $k$ and
$1^{T(\vx)}$ in tape $k+1$ and empty elsewhere, reverses the
computation of $\tilde{M}_{T,k+2}$ in time $p(T(\vx))$.
\end{theorem}

\begin{proofqed}
Let $M=(Q,\{q_0\},\{q_f\},\Sigma^k,\delta)$ be a well-formed QTM. By
Lemma \ref{lemma:final-state}, we have a $(k+1)$-tape, well-formed,
synchronous QTM $M_1$ running in time $2T(\vx)+2$ on input
$(\vx,1^{T(\vx)})$ that satisfies $M_1(\qbit{\vx,1^{T(\vx)}}) =
\tilde{M}_{T,1}(\qbit{\vx,1^{T(\vx)}})$.

By modifying the proof of Proposition \ref{prop:dynamic}, we can show
the existence of a $2(k+1)$-tape, well-formed, stationary,
synchronous, unidirectional, dynamic QTM $M_2$ in normal form such
that (i) $M_2$ on input $(\vx,1^{T(\vx)})$ halts in time
$O(T(\vx)^2)$, (ii) when $M_2$ halts, tape $k+1$ consists only of its
input $1^{T(\vx)}$ and tapes $k+2$ to $2k+2$ are empty, and (iii)
$M_2(\qbit{\vx,1^{T(\vx)}})$ is identical to
$M_1(\qbit{\vx,1^{T(\vx)}})$ when tapes $k+2$ to $2k+2$ are ignored.

It is easy to extend Reversal Lemma in \cite{BV97} to any multi-tape
QTM. Let $M^R$ be the QTM (as constructed in \cite{BV97}) that
reverses the computation of $M_2$ with extra two steps.  Since $M_2$
is well-formed, synchronous, and dynamic, so becomes $M^R$ because of
its construction.  Since any final superposition of $M_2$ is identical
to that of $\tilde{M}_{T,k+2}$, the theorem follows.
\end{proofqed}

Theorem \ref{theorem:reverse} leads to the following lemma.  The proof
of the lemma also uses an argument similar to that of Theorem 4.14 in
\cite{BBBV97}.

\begin{lemma}\label{lemma:squaring}(Squaring Lemma)\hs{2}
Let $k\geq2$. Let $M$ be a $k$-tape, well-formed QTM with a single
final state which, on input $\vx$, outputs $b(\vx)\in\{0,1\}$ in the
start cell of tape $k$ in time $T(\vx)$ with probability
$\rho(\vx)$. There exist a quadratic polynomial $p$ and a
$(2k+3)$-tape, well-formed, stationary, normal form QTM $M'$ such
that, on input $(1^{T(\vx)},\vx)$, $M'$ reaches in time $p(T(\vx))$
the configuration in which $M'$ is in a single final state with
$1^{T(\vx)}$ in tape $1$, $\vx$ in tapes $2$ to $k$, $b(\vx)$ in tape
$k+1$, and empty elsewhere, with probability $\rho(\vx)^2$.
\end{lemma}

\begin{proofqed}
Let $M$ be a given QTM. By Theorem \ref{theorem:reverse}, there exists
a $2(k+1)$-tape, well-formed, synchronous, dynamic, normal form QTM
$M^R$ that, on input $(1^{T(\vx)},\vx)$, reverses the computation of
$\tilde{M}_{T,k+2}$ in time $O(T(\vx)^2)$.

We define the desired QTM $M'$ as follows. Let $(1^{T(\vx)},\vx)$ be
any input. Starting with its initial configuration $cf_0$, $M'$ runs
$\tilde{M}_{T,k+2}$ with ignoring tape $2k+3$. Consider the final
superposition $\tilde{M}_{T,k+2}(\qbit{1^{T(\vx)},\vx})$. When
$\tilde{M}_{T,k+2}$ halts, $M'$ copies the content of the start cell
in the output tape into tape $2k+3$ in two extra steps. Now we have
the superposition
$\qbit{\phi}=\sum_{y}\alpha_{\svx,y}\qbit{y}\qbit{b_y}$, where
$b_y\in\{0,1\}$ is the content of tape $2k+3$ and $y$ ranges over all
configurations excluding the status of tape $2k+3$.  Next, $M'$ runs
$M^R$ starting with $\qbit{\phi}$ with ignoring tape $2k+3$.  Note
that $M^R(\qbit{\phi'})=\qbit{cf_0}\qbit{b(\vx)}$ for the
superposition
$\qbit{\phi'}=\sum_{y}\alpha_{\svx,y}\qbit{y}\qbit{b(\vx)}$.

By a simple calculation, we have $\measure{\phi'}{\phi}=
\sum_{y:b_y=b(\svx)}|\alpha_{\svx,y}|^2$, which equals $\rho(\vx)$
since $\tilde{M}_{T,2k+3}$ outputs $b(\vx)$ with probability
$\rho(\vx)$. 

Since $M^R$ preserves the inner product,
$\measure{M^R(\qbit{\phi'})}{M^R(\qbit{\phi})}=
\measure{\phi'}{\phi}$, which is the amplitude in
$M'(\qbit{1^{T(\vx)},\vx})$ of $\qbit{cf_0}\qbit{b(\vx)}$. Thus, the
squared magnitude of amplitude of $\qbit{cf_0}\qbit{b(\vx)}$ is
exactly $\rho(\vx)^2$.
\end{proofqed}

\paragraph{\bf Timing Problem.}\label{sec:timing}

Let $M=(Q,\{q_0\},\{q_f\},\Sigma^k,\delta)$ be a $k$-tape, partial,
well-formed, normal form QTM. We assume that any computation path of
$M$ on input $\vx$ is completely determined by $\delta$ and ends with
final state $q_f$ and that the length of any computation path of $M$
on $x$ does not exceed $T(\vx)$. We modify $\delta$ by forcing
$\delta(q_f,\vsigma)$ to be $\qbit{q_f}\qbit{\vsigma}\qbit{\vN}$ for
any $\vsigma\in\Sigma^k$ and let $\delta_*$ denote this modified
$\delta$. This $\delta_{*}$ makes $M$ halt within time $T(\vx)$. For
clarity, let $M_*$ be the QTM defined by $\delta_{*}$. Although $M_*$
may not be well-formed, when the final superposition has unit
$L_2$-norm, we can still consider the acceptance probability of $M_*$
as before. Can we simulate $M_*$ on a well-formed QTM?

For convenience, we say that $M$ is {\em well-structured} if (1) it is
well-formed, (2) any computation path of $M$ on input $x$ is
completely determined by $\delta$ and ends with a single final state,
and (3) any final superposition of $M_*$ on each input has unit
$L_2$-norm. For simplicity, we write $\mu_M(\vx)$ to denote the
acceptance probability of $M_*$ on input $\vx$.

\begin{lemma}\label{lemma:timing}
Let $M$ be a $k$-tape, well-structured, partial QTM in normal form
such that the length of any computation path of $M$ on each input
$\vx$ is less than $T(\vx)$. There exists a $(k+3)$-tape, well-formed
QTM $M'$ such that, on input $\vx$ in tapes $1$ to $k$ and
$1^{T(\vx)}$ in tape $k+1$ and empty elsewhere, it halts in time
$O(T(\vx)^2)$ and satisfies $\mu_{M'}(\vx,1^{T(\vx)})=
2^{-\floors{\log T(\vx)}-1}\mu_{M}(\vx)$.
\end{lemma}

\begin{proofqed}
Let $M$ be a given QTM. We first construct a well-formed QTM $M_1$
running in time $O(T(\vx)^2)$ on input $(\vx,1^{T(\vx)})$ such that
the probability that $M$ halts in an accepting configuration in which
tape $k+2$ consists only of symbols $0^{\floors{\log T(\vx)}+2}$, is
$2^{-\floors{ \log T(\vx)}-1}\mu_{M}(\vx)$.

1. We produce in tape $k+3$ the ``reversed'' binary representation of
$T(\vx)$ in exactly $2T(\vx)^2+12T(\vx)+9$ steps. Using this
representation, we produce $\floors{\log T(x)}+1$ bit zeros (following
control bit $1$) in tape $k+2$.

2. We simulate $M$'s move by incrementing two {\em counters}. The
   first counter is in tape $k+1$, of unary form, and the second one
   is a binary counter in tape $k+2$. At each round of simulating a
   single step of $M$, $M_1$ also increments the unary counter by
   stepping right and increments the binary one (using control bit
   $1$) in exactly $2\floors{\log T(\vx)}+8$ steps. When $M$
   terminates, $M_1$ keeps incrementing the unary counter but idles on
   the binary counter for each $2\floors{\log T(\vx)}+8$ steps (using
   control bit $0$ in tape $k+2$ for reversibility).

3. After $T(\vx)$ rounds, we apply a Hadamard transform, with stepping
right, to the content of the binary counter except its control bit
(\ie $\delta'(p,\vsigma*\sigma')=
\frac{1}{\sqrt{2}}\sum_{\tau\in\{0,1\}} (-1)^{\sigma'\cdot\tau}
\qbit{p}\qbit{\vsigma,\tau}\qbit{\vN,R}$, where $\sigma'$ is in tape
$k+2$). Since the length of this counter is $\floors{\log T(\vx)}+1$,
we can observe symbols $0\circ 0^{\floors{\log T(\vx)}+1}$ in tape
$k+2$ with amplitude $2^{-\floors{ \log T(\vx)}-1}$. Hence, the
probability that $M$ reaches an accepting configuration with
$0^{\floors{\log T(\vx)}+2}$ in tape $k+2$ is $2^{-\floors{ \log
T(\vx)}-1}\mu_{M}(\vx)$.

We design $M'$ so that the heads in tapes $k+1$ to $k+3$ return to the
start cells (using $1^{T(\vx)}$ in tape $k+1$) and the rest of heads
stay in the same cells as $M$'s. It is easy to see that $M_1$ is in
normal form if we add the rule:
$\delta'(q_f,\vsigma)=\qbit{q_0}\qbit{\vsigma}\qbit{\vN}$. Moreover,
if $M$ is stationary, $M_1$ is also stationary.

For the desired machine $M'$, we design it to accept input
$(\vx,1^{T(\vx)})$ exactly when $M_1$ reaches an accepting
configuration with $0^{\floors{\log T(\vx)}+2}$ written in tape $k+2$.
It thus follows that $\mu_{M'}(\vx,1^{T(\vx)}) = 2^{-\floors{\log
T(\vx)}-1}\mu_{M}(\vx)$.
\end{proofqed}

Lemma \ref{lemma:timing} solves the timing problem for any quantum
complexity class whose acceptance criteria is invariant to a
polynomial fraction of acceptance probability.


\section{Oracle Quantum Turing Machines}\label{sec:oracle}

Unlike the previous sections, we will focus on an oracle QTM, which is
a natural extension of a classical oracle TM with the help of a set of
oracles.

Formally, we define a $(k+m)$-tape {\em oracle QTM $M$ with $m$ query
tapes} to be a septuple $(Q,\{q_0\},Q_f,
Q_p,Q_a,\Sigma_1\times\Sigma_2\times\cdots\times\Sigma_{k+m},\delta)$,
where $Q$ includes $Q_p=\{q_p^1,q_p^2,\ldots,q_p^m\}$, a set of
pre-query states, and $Q_a=\{q_a^1,q_a^2,\ldots,q_a^m\}$, a set of
post-query states, and the transition function $\delta$ is defined
only on $(Q-Q_p)\times\Sigma^k$. We assume the reader's familiarity
with an {\em oracle query}. For its definition, see
\cite{BV97}. Conventionally, we assume that every alphabet
$\Sigma_{k+i}$, $1\leq i\leq m$, includes binary bits $\{0,1\}$.  Let
$\mathcal{A}=(A_i)_{1\leq i\leq m}$ be a series of oracles such that
each $A_i$ is a subset of $(\Sigma_{k+i})^*$.  Note that query states
$q_p^i$ and $q_a^i$ correspond only to the $i$th query tape and the
$i$th oracle $A_i$.

It is important to note that Well-Formedness Lemma and Completion
Lemma hold even for oracle QTMs.

\paragraph{\bf Reducing the Number of Query Tapes.}

We can reduce the number of query tapes by combining a given set of
oracles into a single oracle together with copying a query word
written in one of query tapes into a single query tape. When we copy a
query word $y\circ b$ from the $i$th query tape, we pad the suffix
$0^{i}1^{m-i}$ (between $y$ and $b$) to make the copying process
reversible.

\begin{lemma}\label{lemma:number-states}
Let $m\geq2$. Let $M$ be a $(k+m)$-tape, well-formed, oracle QTM with
$m$ query tapes that halts in time $T(\vx)$ on input
$\vx\in\Sigma^k$. Let $\mathcal{A}=(A_i)_{1\leq i\leq m}$ be a series
of oracles. There exists a $(k+2m+1)$-tape, well-formed, oracle QTM
$M'$ with a single query tape such that, on input $(\vx,1^{T(\vx)})$,
halts in time $5T(\vx)^2+8T(\vx)$ and $\mu_{M'}^{B}(\vx,1^{T(\vx)})=
\mu_M^{\mathcal{A}}(\vx)$, where $B=\{y0^i1^{m-i}\mid y\in A_i\}$.
\end{lemma}

\paragraph{\bf Adjusting the Number of Queries.}

Let $M$ be a given QTM. At the end of each round, in which a new QTM
$M'$ simulates a single step of the computation of $M$, we force $M'$
to make a query (of the form $0\circ 0$) in 6 steps if $M$ does not
query.  When $M$ invokes an oracle query, we force $M'$ to idle for 6
steps instead of making a query of $0\circ 0$. This proves the lemma
below.

\begin{lemma}\label{lemma:query-number}
Let $M$ be a $(k+1)$-tape, well-formed, oracle QTM in normal form with
a single query tape that halts in time $T(\vx)$d, on input
$\vx\in\Sigma^k$. Let $A$ be an oracle. There exist a $(k+2)$-tape,
well-formed, oracle QTM $M'$ with two query tapes running in time
$7T(\vx)$ on input $\vx$ such that $M'$ makes exactly $T(\vx)$ queries
along each computation path and $\mu_{M'}^{(A,A)}(\vx) =
\mu_M^A(\vx)$.
\end{lemma}

\paragraph{\bf Adjusting the Length of Query Words.}

We show that the length of query words can be stretched with quadratic
slowdown. To extend the length of a query word to the fixed length
$T-1$, we pad the suffix $01^{T-|y|-2}$ in $4T+6$ steps.

\begin{lemma}\label{lemma:query-length}
Let $M$ be a $(k+1)$-tape, well-formed, oracle QTM in normal form with
a single query tape that halts in time $T(\vx)$ on input
$\vx\in\Sigma^k$. Let $A$ be an oracle set. There exists a
$(k+3)$-tape, well-formed, oracle QTM $M'$ such that, for every input
$(\vx,1^{T(\vx)})$, it halts in time $4T(\vx)^2+10T(\vx)$, the length
of any query word is exactly $T(\vx)-1$ on any computation path, and
it satisfies $\mu_{M'}^B(\vx,1^{T(\vx)}) = \mu_M^A(\vx)$, where
$B=\{y01^{m-|y|-2}\mid y\in A,m\geq|y|+2\}$.
\end{lemma}
\ms

\centerline{\large {\bf Acknowledgements}}\bs

The author is grateful to Andy Yao and Yaoyun Shi for interesting
discussion o quantum computation.

\bibliographystyle{alpha}

\end{document}